\documentstyle[12pt,epsf,here,rotate]{article}
\begin{document}
\title{{
Broken Pairs and High-Spin States\\
in Transitional Nuclei}}
\author{{ } 
G. Bonsignori, D. Vretenar, S. Cacciamani and L. Corradini\\
INFN and Physics Department, University of Bologna, Italy}
\date{}
\maketitle
\begin{abstract}
The Interacting Boson plus broken-pairs Model has been used to
describe high-spin bands in spherical and transitional nuclei.
In the spherical nucleus $^{104}$Cd the model reproduces the structure
of high-spin bands built on a vibrational structure. Model calculations
performed with a single set of parameters reproduce ten bands in $^{136}$Nd,
including the two dipole high-spin bands based on the  
$(\pi h_{11/2})^2$ $(\nu h_{11/2})^2$ configuration.
\end{abstract} 


\section{Introduction}
Models that are based on the Interacting Boson Approximation provide a 
consistent description of low-spin nuclear structure in spherical, deformed 
and transitional nuclei. By including selective non-collective
fermion degrees of freedom,
through successive breaking of correlated S an D pairs, the IBM can be
extended to describe the physics of high-spin states in nuclei.
This extension of the model is especially relevant for transitional regions, 
where single-particle excitations and vibrational collectivity are dominant 
modes and no pronounced axis for cranking exists. We present a model that
extends the IBM-1 to 
include two- and four-fermion noncollective 
states (one and two broken-pairs).
The model has been applied in the description of 
high-spin states in the Hg~\cite{IV91,VRE93,VRE95}, 
Sr-Zr~\cite{CHO91,LIS93,CHI93,caccia},
and Nd-Sm~\cite{DEA94,RAV96,PET97} regions. 

The model is based on the IBM-1 \cite{IAC87};
the boson space consists of {\it s} and {\it d} bosons, with no distinction
between protons and neutrons. To generate high-spin states,
the model allows one or two bosons to be destroyed and non-collective
fermion pairs formed, represented by two- and four-quasiparticle
states which recouple to the boson core.
High-spin states are described in terms of broken pairs.
The model space for an even-even nucleus with $2N$ valence nucleons is
\begin{eqnarray}
\mid N~bosons~> & \oplus & \mid (N-1)bosons \otimes 1~broken~pair> \nonumber \\
& \oplus & \mid (N-2)bosons \otimes 2~broken~pairs> \oplus~~... \nonumber
\end{eqnarray}
The model Hamiltonian has four terms:
the IBM-1 boson Hamiltonian, the fermion Hamiltonian, the
boson-fermion interaction, and a pair breaking interaction that
mixes states with different number of fermions.
\begin{equation}
H=H_{B}+H_{F}+V_{BF}+V_{mix}.
\end{equation}
The fermion Hamiltonian $H_F$ contains single-fermion energies and 
fermion-fermion interactions. 
The interaction between the unpaired fermions
and the boson core contains the dynamical, exchange and
monopole interactions of the IBFM-1~\cite{IAC79} 
In order to describe dipole bands in transitional nuclei, we have modified
the quadrupole-quadrupole dynamical interaction 
\begin{equation}
V_{dyn}=\Gamma _{0}\sum_{j_{1}j_{2}}(u_{j_{1}}u_{j_{2}}-v_{j_{1}}v_{j_{2}})
\langle j_{1}\parallel Y_{2}\parallel j_{2}\rangle \times 
\left( [a^{\dagger}_{j_{1}} \times\tilde{a}_{j_{2}}]^{(2)}
\cdot Q^{B} \right),
\end{equation}
The standard boson quadrupole operator $Q^{B}$ has been 
extended by a higher order term
\begin{equation}
               \chi'\sum_{L_{1}L_{2}}\left[
                                     \left[
                                           d^{\dag} \times \tilde{d}
                                     \right]^{\left( L_{1} \right)}
                                     \times
                                     \left[
                                           d^{\dag} \times \tilde{d}
                                     \right]^{\left( L_{2} \right)}
                                     \right]^{\left( 2 \right)}\;,
\label{chi_prime}
\end{equation}
The terms $H_B$, $H_F$ and $V_{BF}$ conserve the
number of bosons and the number of fermions separately. In our
model only the total number of nucleons is conserved, bosons
can be destroyed and fermion pairs created, and vice versa. In
the same order of approximation as for $V_{BF}$,
the pair breaking
interaction $V_{mix}$ which mixes states with different number of
fermions, conserving the total nucleon number only, reads
\begin{eqnarray}
V_{mix}&=&-U_{0}\left\{ \sum_{j_{1}j_{2}}u_{j_{1}}u_{j_{2}}(u_{j_{1}}v_{j_{2}}+
u_{j_{2}}v_{j_{1}})\langle j_{1}\parallel Y_{2}\parallel j_{2}\rangle ^{2}
\frac{1}{\sqrt{2j_{2}+1}}\left( [a^{\dagger}_{j_{2}}\times
a^{\dagger}_{j_{2}}]^{(0)}\cdot s\right)+\ h.c.\right\}  \nonumber\\
&&-U_{2}\left\{ \sum_{j_{1}j_{2}}(u_{j_{1}}v_{j_{2}}+
u_{j_{2}}v_{j_{1}})\langle j_{1}\parallel Y_{2}\parallel j_{2}\rangle
\left( [a^{\dagger}_{j_{1}}\times a^{\dagger}_{j_{2}}]^{(2)}
\cdot \tilde{d}\right) +\ h.c.\right\}.
\label{mixing}
\end{eqnarray}
If mixed proton-neutron configurations are included in the 
fermion model space, i.e. there can be both proton and 
neutron broken pairs, the full model Hamiltonian reads
\begin{equation}
                H = H_{B} + H_{\nu F} + H_{\pi F} + H_{\nu BF}+
                 H_{\pi BF} + H_{\nu}^{mix} + H_{\pi}^{mix} + H_{\nu \pi},
\end{equation}
where the proton-neutron interaction term is defined
\begin{eqnarray}
             H_{\nu \pi} & = & \sum_{nn'pp'} \sum_{J} (-)^{J}
                                       h_{J}(nn'pp')
                                      (u_{n}u_{n'}-v_{n}v_{n'})
                                        (u_{p}u_{p'}-v_{p}v_{p'})
                                        \times \nonumber  \\
                                  &   & \times \left(
                                               \left[
                                 a^{\dag}_{n} \times \tilde{a}_{n'}
                                               \right]^{(J)} \cdot
                                               \left[
                                     a^{\dag}_{p} \times \tilde{a}_{p'}
                                               \right]^{(J)}
                                               \right)\;.
\end{eqnarray}
In the following two sections we present results
for quadrupole bands in the spherical nucleus
$^{104}$Cd, and for proton-neutron
dipole bands in the transitional nucleus $^{136}$Nd.

\section{ The nucleus $^{104}_{~48}$Cd$_{56}$ }
The Cd isotopes, with only two protons from the closed shell at
N=50, present an example of spherical vibrational nuclei 
in which also high-spin quadrupole bands are found.
In particular, in the nucleus $^{104}$Cd quadrupole bands extend  
up to spin 26 $\hbar$. Here we present a
description of the structure of the excitation
spectrum in the framework of the IBM plus one broken pair.

In general most of the parameters of the Hamiltonian are taken
from analyses of
the low- and high-spin states in the neighboring even and odd nuclei. 
For $^{104}$Cd the parameters of the boson Hamiltonian are:
$\epsilon=0.658$ MeV, $C_4=0.117$ MeV, the number of bosons is $N=4$. 
$\epsilon$ corresponds to the excitation energy of $2^+_1$, and $C_4$ is 
adjusted to reproduce the $6^+_1$ and $8^+_1$ states. 
Since only the yrast sequence of the collective vibrational 
structure is known 
experimentally, the remaining parameters of the 
boson Hamiltonian could not be 
determined, and are set to zero. 

The resulting SU(5) vibrator spectrum displays very little anharmonicity.
In the present calculation we only consider collective states and
two-quasiparticle states based on configurations with two neutrons in the 
broken pair. States based on the proton $(g_{9/2})^2$ configuration are not 
included in the model space.
The single-quasiparticle neutron energies and occupation probabilities are 
obtained by a simple BCS calculation.
The quasiparticle
energies and occupation probabilities are:
$E (\nu d_{5/2})=1.113$ MeV,
$E (\nu s_{1/2})=2.287$ MeV,
$E (\nu h_{11/2})=2.691$ MeV,
$E (\nu g_{7/2})=1.316$ MeV,
$v^{2} (\nu d_{5/2})=0.57,$
$v^{2} (\nu s_{1/2})=0.06,$
$v^{2} (\nu h_{11/2})=0.04,$
$v^{2} (\nu g_{7/2})=0.23.$

The parameters of the boson-fermion interactions have been adjusted
to reproduce the lowest positive and negative parity structures in the 
neighboring odd-N isotopes $^{103}$Cd and $^{105}$Cd. 
For neutron states the boson-fermion parameters are: $\Gamma_0=0.2$ MeV and 
$\chi=-0.9$ for the dynamical interaction, and $\Lambda_0=0.2$ MeV for the 
exchange interaction. The parameter of the boson quadrupole operator, 
$\chi=-0.9$, is adjusted to reproduce the experimental data for 
$^{106}$Cd: $B(E2, 2_1 \rightarrow 0_1) = 0.068~e^2b^2$ 
and $Q(2_1)=-0.25~eb$. 
In addition to the boson and boson-fermion parameters that have been 
already discussed, the strength parameter of the pair-breaking interaction 
is $U_2=0.1$ MeV. 
Results of model calculations for positive parity states
in $^{104}$Cd, with a fermion space of one neutron broken pair, are shown 
in Fig. 1.

In this energy level diagram, only few lowest calculated
levels of each spin are compared to the experimental counterparts. 
The collective vibrational structure remains yrast up to angular momentum 
$I=6^+$. In the experimental spectrum one finds, below the collective 
$I=8^+$ (3211 keV) state, two other states of the same angular momentum: 
at 2902 keV and 3031 keV. They are probably based on a two-neutron 
configurations. 
It is interesting that this triplet of $I=8^+$ states is also 
reproduced by our calculation. In fact, for three 
$\Delta I=2$ positive parity sequences, based on the $(d_{5/2})^2$, 
$(g_{7/2})^2$ and $(d_{5/2}, g_{7/2})$ neutron configurations,
probable experimental counterparts are observed. 
The calculated energy spacings correspond to the collective vibrational 
sequence. The experimental energy spacings are somewhat smaller,
indicating a stronger core polarization and/or change of 
deformation. This is probably caused by admixtures of 2p-2h proton 
configurations, an effect that could not be included in our model space. 
In heavier Cd isotopes experimental data exists on the neutron $(h_{11/2})^2$
structure. The $I=10^+$ band-head of the $\Delta I=2$ $(h_{11/2})^2$ sequence 
is at: 4153 KeV in $^{108}$Cd, and 4816 keV in $^{106}$Cd. In $^{104}$Cd this 
state is not observed, our calculations place it at 5310 keV. $^{104}$Cd 
has also less particles outside the closed shell compared 
to heavier isotopes,
and therefore collective properties are less pronounced.
In particular, with only four bosons we can only construct 
states up to angular 
momentum $I=16^+$, including the fermion space of two neutrons. States with 
higher angular momenta should be based on a different core, probably 
one that includes proton excitations across the $Z=50$ shell closure.
 
The lowest calculated negative-parity two neutron states are compared with
experimental levels in Fig. 2. 
The parameters of the Hamiltonian have the same values as in the calculation 
of positive-parity states. 
There are several negative parity sequences in the experimental 
spectrum which start at $\approx 4$ MeV, and with angular momenta 
$I=7^-~-~9^-$. The lowest two calculated levels of each spin $I\geq 7^-$
are displayed in Fig. 2. The levels are grouped into 
sequences according to the dominant fermion component in the 
wave function. The two-fermion angular momentum is approximately 
a good quantum number. The collective part of the wave 
functions correspond to that of the low-lying collective vibrational
sequence. Because the quasiparticle energies of $d_{5/2}$ and $g_{7/2}$
differ by only $\approx 200$ keV, both the $(h_{11/2},d_{5/2})$ and
the $(h_{11/2},g_{7/2})$ configurations form sequences with 
states of the same angular momentum at almost the same excitation
energies. Of course the density of negative parity states 
between 4 MeV and 7 MeV is very high. All four neutron 
orbitals are included in the calculation, and one finds 
many states with low angular momenta, that are not observed 
in experiment. Therefore in Fig. 2 we only display
states for which correspondence with experimental data
can be established.
\section{ The nucleus $^{136}_{~60}$Nd$_{76}$}
Nuclei in the A = 130-140 mass region are $\gamma$-soft and the 
polarizing effect of the aligned nucleons
induces changes in the nuclear shape.
Because of the different nature of the excitations (particles for 
proton, and holes for neutron configurations), the alignment of a pair of
$h_{11/2}$ protons induces a prolate shape, whereas the alignment 
of a neutron pair in the $h_{11/2}$ orbital 
drives the nucleus towards a collective oblate shape. In $^{136}$Nd
one therefore expects to observe different 
coexisting structures at similar excitation energies.
The IBM model with broken pairs has been previously 
used in the description of
low-spin and high-spin properties of $\gamma$-soft
nuclei of this region (in the IBM language O(6) nuclei):
$^{138}$Nd~\cite{DEA94}, $^{137}$Nd ~\cite{PET97} and 
$^{139}$Sm~\cite{RAV96}.

In the present work we use the IBM with proton and neutron broken 
pairs to describe the excitation spectrum of $^{136}$Nd. In 
particular, we want to obtain a correct description 
of high-spin dipole bands, which have been interpreted 
as two proton - two neutron structures.
The experimental level scheme of positive-parity states
is displayed in Fig. 3. The labels of bands are from 
Ref.~\cite{SUN96}; in addition to the ground state band 
and the quasi $\gamma$-band, bands 3, 5, 7 and 8 
result from the alignment of two protons or two 
neutrons in the $h_{11/2}$ orbital, the two 
four-quasiparticle dipole bands have labels 10 and 11.
 
There are 6 neutron valence {\em holes} and 10 proton 
valence {\em particles}. The resulting boson number is N=8.
The set of parameters for the boson Hamiltonian is:
$\epsilon$=0.36, $C_0$=0.16, $C_2$=-0.12, $C_4$=0.19,
$V_2$=0.11 and $V_0$=-0.3 (all values in MeV).
The boson parameters have values similar to those that
have been used in the calculation of $^{138}$Nd, 
$^{137}$Nd and $^{139}$Sm.

In A $\approx$ 140 nuclei
the structure of positive parity high-spin states close to the 
yrast line is characterized by the alignment of both 
proton and neutron pairs in the $h_{11/2}$ orbital. 
For positive-parity states we have only included the 
proton and neutron $h_{11/2}$ orbitals in the fermion
model space. Additional single-nucleon states 
make the two broken-pairs bases prohibitively large. 
The single quasiparticle energies and occupation probabilities are obtained 
from a  BCS calculation. 
Similar to our previous calculations 
for $^{138}$Nd and $^{137}$Nd, the resulting quasiparticle energies 
for the proton and neutron $h_{11/2}$ states had to be slightly
renormalized. 
$E_{\nu}(h_{11/2}) = 1.75$ MeV, $v^2_{\nu}(h_{11/2}) = 0.83$, 
$E_{\pi}(h_{11/2}) = 1.60$ MeV, $v^2_{\pi}(h_{11/2}) = 0.07$.
In order to further reduce the large size of the 
space with two-broken pairs, we had prediagonalized the 
boson Hamiltonian, and the fermion states were then coupled 
to the lowest eigenvectors, i.e. only to the collective 
ground state band.
The parameters of the fermion-boson interactions are determined from IBFM 
calculations of low-lying negative-parity states in $^{137}$Nd and 
neighboring odd-proton nuclei.
The parameters of the neutron dynamical fermion-boson interaction are
$\Gamma_0$=0.3 MeV, $\chi$=-1 and $\chi'$=-0.2, and for protons: 
$\Gamma_0$= 0.22 MeV, $\chi$=+1 and $\chi'$=+0.2.
For the exchange interaction $\Lambda_0^{\nu}=1.0$ and
$\Lambda_0^{\pi}=1.5$ for neutrons and protons, respectively.
The strength parameter of the pair-breaking interaction is $U_0=0$ and
$U_2=0.2$ MeV, both for protons and neutrons in broken pairs.
The residual interaction between unpaired fermions is a surface 
$\delta$-force with strength parameters:
$ V_{\nu\nu} = -0.1 $, $ V_{\pi\pi} = -0.1$ and
$ V_{\nu\pi} = -0.9 $ for neutron-neutron, proton-proton and 
proton-neutron, respectively.

In Fig. 4 we display the calculated spectrum of positive-parity 
states. According to the structure of wave functions, states 
are classified in bands labeled in such a way that a direct
comparison can be made with their experimental counterparts.
The calculated positive-parity structures 
3, 5, 7, 8, 10 and 11, as well as the ground state band an the 
quasi $\gamma$-band, have to be compared with the experimental bands 
of Fig. 3. The bands 3, 5, and 7 result from the alignment
of a pair of protons in the $h_{11/2}$ orbital. The band
8 corresponds to two $h_{11/2}$ neutrons coupled to the boson core. 
Finally, the two dipole bands 10 and 11 correspond to four-quasiparticle
states, two protons and two neutrons in their respective 
$h_{11/2}$ orbitals, coupled to the ground state band of the core.
The occurrence of regular dipole bands ($\Delta I=1$) in nearly 
spherical and transitional nuclei presents an interesting phenomenon. 
In the semiclassical picture of the cranked shell model,
$\Delta I=1$ high-spin bands have been described 
as TAC (Tilted Axis Cranking) solutions~\cite{fra93}. 
In our model such $\Delta I=1$ structures are produced by the
fermion-boson interactions. However, in order to obtain
the correct energy spacings for bands 10 and 11, it was necessary
to include the additional term~(\ref{chi_prime}) in the 
boson quadrupole operator. We have also found that a crucial    
role in the excitation spectrum of these bands is played by 
the proton-neutron delta-interaction. 

It should be noted that
bands 10 and 11 have been recently described in the 
framework of the projected shell model~\cite{SUN96}. In order 
to obtain bands of dipole character, a permanent deformation
had to be assumed, which in turn made energetically more favorable
one of the neutrons to occupy the $\nu f_{7/2}$ orbital. Therefore
the configuration $(\pi h_{11/2})^2$ $\nu f_{7/2}$ $\nu  h_{11/2}$
was assigned to the bands 10 and 11.
The dipole character of the bands results from the coupling 
of the neutron hole in $h_{11/2}$ and the neutron particle
in $f_{7/2}$. 
We believe that this interpretation is not correct. In 
our calculations, structures based on the $\nu f_{7/2}$ orbital
are found high above the yrast. Using a unique set of 
parameters, we have been able to reproduce the complete 
experimental spectrum of positive parity states, from the 
ground state band, to angular momentum 29 $\hbar$ in band
11, at more than 13 MeV excitation energy. 
With the same set of parameters 
we have also calculated 
negative parity states based on the neutron 
orbitals $s_{1/2}$, $d_{3/2}$ and $h_{11/2}$. The resulting
bands reproduce the experimental data.
\subsection{Conclusions}
We have used an extension of the Interacting Boson Model to describe the
physics of high-spin states in nuclei.
Compared with traditional models based on the cranking approximation, 
the present approach provides several advantages. High-spin bands can be 
described not only in well deformed,
but also in transitional and spherical nuclei. A single set of parameters
and a well defined Hamiltonian are used to calculate both 
ground state collective bands and high-spin two- and 
four-quasiparticle bands. 
Polarization effects directly 
result from the model boson-fermion interactions.  
The model produces not only energy spectra, but also wave functions which 
can be used to calculate electromagnetic properties. 
All calculations are performed in the laboratory frame, and therefore  
produce results that can be 
directly compared with experimental data.

In the spherical nucleus $^{104}$Cd the model reproduces the structure
of high-spin bands built on a vibrational structure. The bands result
from the alignment of a pair of neutrons in the orbitals $\nu d_{5/2}$
and $\nu g_{7/2}$. For the transitional nucleus $^{136}$Nd, our 
calculation reproduce the complete experimental excitation 
spectrum of positive and negative parity states. In particular,
we have been able to obtain a correct description of the 
two high-spin $(\pi h_{11/2})^2$ $(\nu h_{11/2})^2$ dipole bands.


\vspace{2 cm}
\leftline{\bf Figure Captions}
\begin{description}
\item {\small {\bf FIG. 1.}
Results of the IBM plus broken pair calculation
for positive-parity states compared with experimental
levels in $^{104}$Cd.}
\item {\small {\bf FIG. 2.}
Negative-parity states in $^{104}$Cd compared with results
of the IBM plus broken pair calculation.}
\item {\small {\bf FIG. 3.}
Experimental excitation spectrum of positive-parity 
states in $^{136}$Nd.}
\item {\small {\bf FIG. 4.}
Results of IBM plus broken pairs calculation for positive
parity bands in $^{136}$Nd}
\end{description}
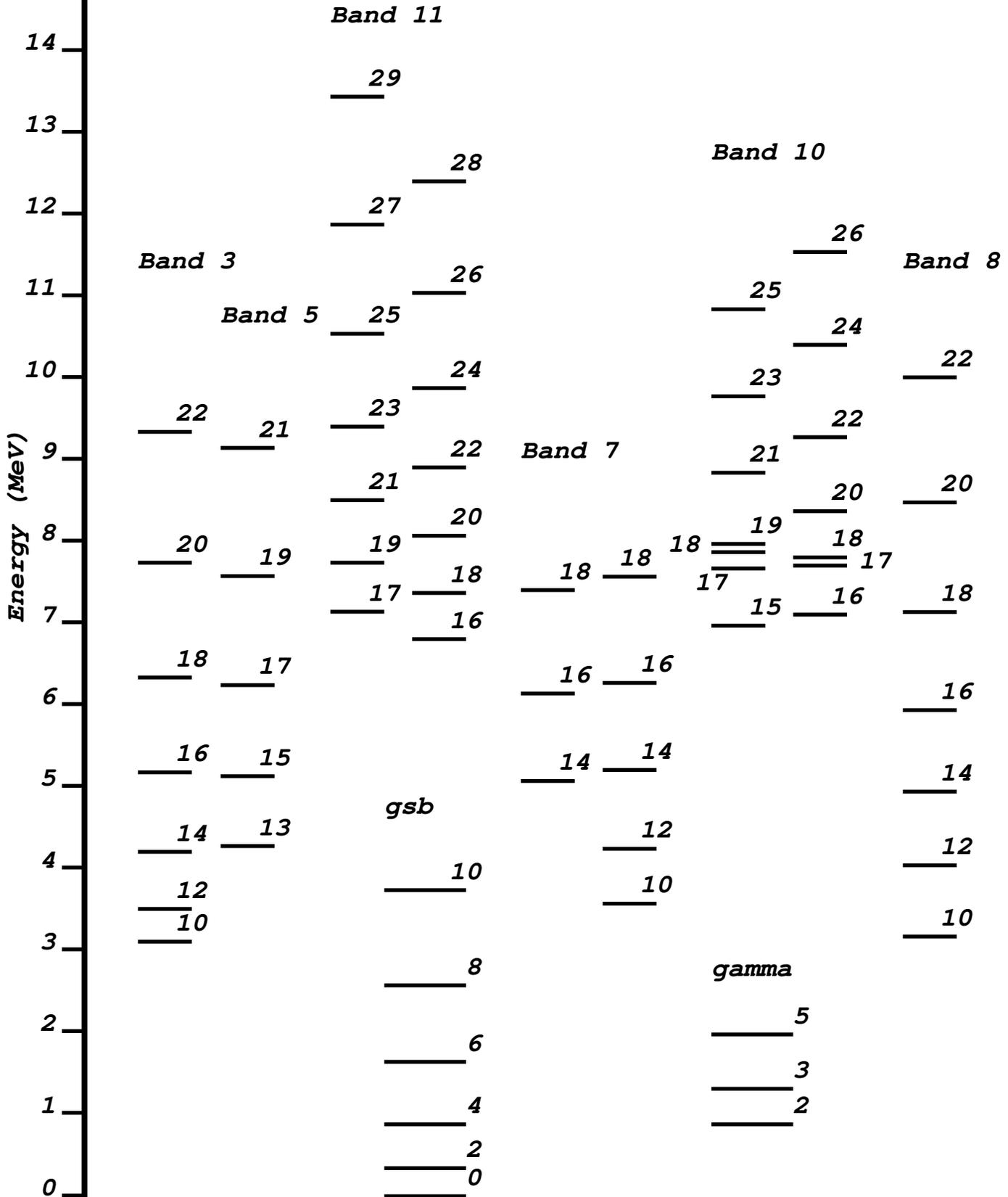
\end{document}